\documentclass[aps,prd,nofootinbib,preprint]{revtex4}
\usepackage{graphicx}
\usepackage{psfrag}

\begin{document}
\title{Hadronic production of $B^{(*)}_s$ at TEVATRON and LHC}
\author{Jia-Wei Zhang$^{1}$, Zhen-Yun Fang$^{1}$, Chao-Hsi Chang$^{1,2}$, Xing-Gang Wu$^{1}$\footnote{Email:
wuxg@cqu.edu.cn}, Tao Zhong$^{1}$ and Yao Yu$^{1}$}
\address{$^1$Department of Physics, Chongqing University, Chongqing 400044, P.R.
China\\
$^2$Institute of Theoretical Physics, Chinese Academy of Sciences,
Beijing 100190, P.R. China}

\begin{abstract}
We study the hadronic production of $B_s$ and $B_s^*$ mesons within
the fixed-flavor-number scheme, in which the dominant gluon-gluon
fusion mechanism is dealt with by using the complete $\alpha_s^4$
approach. Main theoretical uncertainties for $B_s$ and $B_s^*$
production at TEVATRON and LHC are presented. It is found that when
$m_s$ increases by steps of $0.1$ GeV, the integrated cross section
of $B^{(*)}_s$ decreases by $80\%-100\%$, when $m_b$ increases by
steps of $0.1$ GeV, it changes to be $\sim 10\%$. While the
uncertainties caused by the parton distribution function and the
factorization scale varies within the region of $~\frac{1}{5}$ to
$~\frac{1}{3}$. Considering possible kinematic cut on the transverse
momentum and the rapidity cut for the detectors at TEVATRON and LHC,
we also make estimations on the $B_s$ and $B_s^*$ production with
various kinematic cuts.  \\


\end{abstract}

\maketitle

Since Run II at the TEVATRON Collider started in 2001, the CDF and
D0 experiments have successfully collected $B_s$ data
\cite{CDF1,CDF2,D0,CDFD0}. One can use $B_s$ meson to study those
interesting topics as QCD model building, physics beyond the
Standard Model, the electro-weak symmetry breaking mechanism,
charge-parity (CP) violation and etc.
\cite{theo1,theo2,theo3,theo4}. Taking into account the prospects of
$B_s$ production at Fermilab TEVATRON and at the newly running CERN
LHC, the future numerous data require more accurate theoretical
predictions, especially on its hadronic production.

According to the QCD factorization formula, the hadronic production
of $B_s$ and $B_s^*$ can be written as
\begin{eqnarray}
d\sigma(S,p_T,\cdots)&=&\sum_{ij}\int\int dx_{1}
dx_{2}F^{i}_{H_1,P_{1}}(x_{1},\mu^2_{F})\cdot
F^{j}_{H_2,P_{2}}(x_{2},\mu^2_{F})\cdot
\nonumber\\
&& d\hat{\sigma}_{ij\rightarrow
B^{(*)}_{s}X}(P_1,P_2,x_{1},x_{2},\mu^2_{F}, Q^2,
\hat{s},p_T,\cdots)\ , \label{cross}
\end{eqnarray}
where $\sqrt{S}$ stands for the total collision energy of the
incoming hadrons, $F^{i}_{H_1,P_{1}}(x_{1},\mu^2_{F})$ and
$F^{j}_{H_2,P_{2}}(x_{2},\mu^2_{F})$ are the parton distribution
functions (PDFs) of incoming hadrons $H_1$ (momentum $P_1$) and
$H_2$ (momentum $P_2$) for parton $i$ (with momentum fraction $x_1$)
and parton $j$ (with momentum fraction $x_2$) respectively. $Q^2$ is
the ``characteristic energy scale of the subprocess squared" and
$\mu_F$ stands for the factorization scale for the PDF and the hard
subprocess. A detailed discussion on the choice of $Q^2$ and $\mu_F$
can be found in Ref.\cite{psi}, here for simplicity, we shall take
$Q^2=\mu^2_F$ for the present perturbative QCD calculation.
$d\hat{\sigma}_{ij\rightarrow B^{(*)}_{s}X}$ stands for the
differential cross-section of the relevant hard subprocess, in which
$\hat{s}=x_1x_2S$ is the c.m.s. energy of the subprocess and $P_T$
is the transverse momentum of $B^{(*)}_s$.

Within the fixed-flavor-number (FFN) scheme \cite{ffn}, where only
light quark/antiquark and gluon should be considered in the initial
state of the hard scattering subprocess, it can be found that
$B^{(*)}_s$ hadronic production are dominated by the gluon-gluon
fusion mechanism, which is through the sub-process $g+g\rightarrow
B^{(*)}_s+b+\bar{s}$ and is of order $\alpha_s^4$. In addition to
the gluon-gluon fusion mechanism, there are several different
mechanisms for the production, such as that via the quark-antiquark
annihilation subprocess $q \bar{q}\to B^{(*)}_s+b+\bar{s}$ and {\it
etc.}. However, it can be found that the contributions to the
production from quark-antiquark annihilation are much smaller (only
about $1\%$) than those from gluon-gluon fusion, which is due to the
fact that the `luminosity' of gluons is much higher than that of
quarks in $pp$ collisions (LHC) and in $p\bar{p}$ collisions
(TEVATRON), and there is a suppression factor due to the virtual
gluon propagator in the annihilation, which is similar to the case
of $B_c$ production \cite{qq0,qq1}. Hence in the present letter, we
shall concentrate our attention on the gluon-gluon fusion mechanism.
And to be useful experimentally, we shall discuss the main
uncertainties in estimating the hadronic production of $B_s^{(*)}$,
which includes the choices of the factorization energy scale
$\mu_F$, the various versions of parton distribution functions
(PDFs), the values of the bound state parameters and etc..

\begin{figure}
\setlength{\belowcaptionskip}{10pt}
\centering
\includegraphics[width=0.70\textwidth]{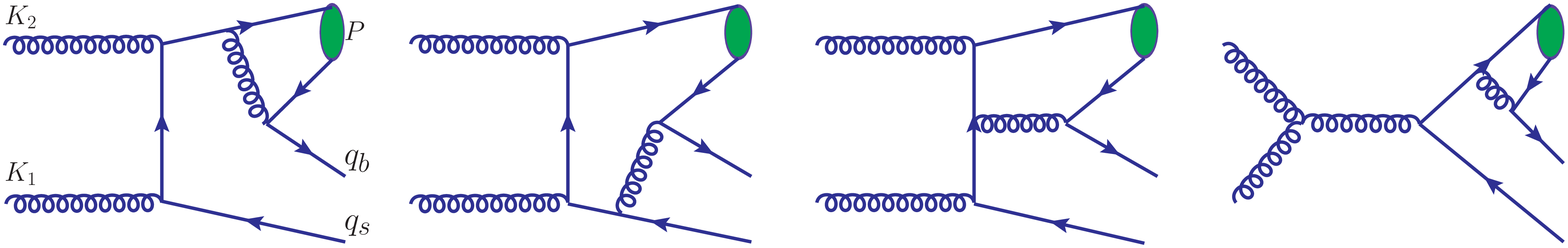}\hfill
\includegraphics[width=0.70\textwidth]{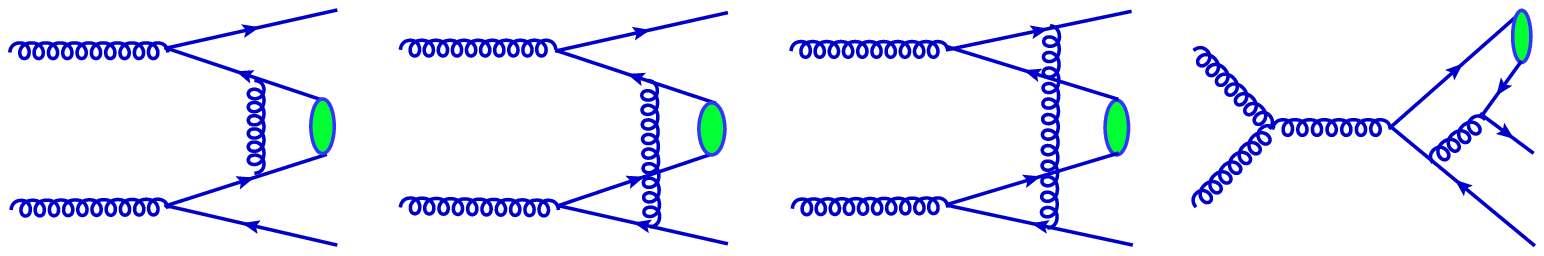}\hfill
\includegraphics[width=0.70\textwidth]{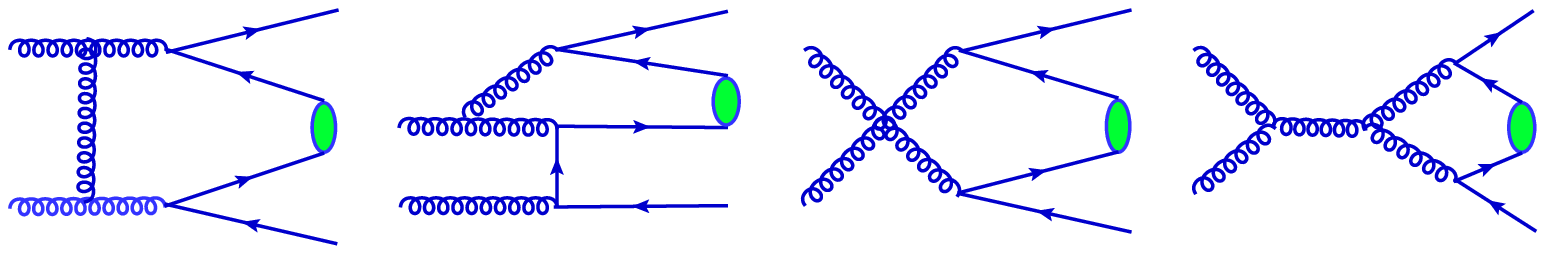}
\caption{Typical Feynman diagrams for the hard subprocess
$g(K_1)+g(K_2)\to B^{(*)}_s(P)+\bar{b}(q_b)+s(q_s)$.} \label{feyn}
\vspace{0mm}
\end{figure}

As for the dominant gluon-gluon fusion mechanism, its hard
subprocess $g+g\to B^{(*)}_s+b+\bar{s}$ includes 36 Feynman
diagrams, whose typical ones are plotted in FIG.\ref{feyn}. To
derive the analytical squared amplitude for this subprocess is a
tedious task, since it contains non-Abelian gluons and massive
fermions. Fortunately however, very recently, a generator BCVEGPY
\cite{BC1,BC2,BC3} for hadronic production of $B_c$ meson has been
available, where to deal with the subprocess $g+g\to B_c
+\bar{b}+c$, the so called `helicity amplitude approach
\cite{hel0,hel1} \footnote{It should be noted that Ref.\cite{hel0}
presented only the formulae for massless spinor lines, so proper
changes as have been done in Ref.\cite{BC1} should be made so as to
deal with the massive spinor lines.} has been adopted to derive
analytic expressions at the amplitude level and then do the
numerical calculation just at the amplitude level. Here we adopt the
same method of Refs.\cite{BC1,BC2,BC3} to deal with the present
$B^{(*)}_s$ production, which can be obtained by suitable changing
the $c$-quark lines defined in Ref.\cite{BC1} to the present
$s$-quark lines. To short the paper, we shall only present the main
idea on how to deal with those 36 Feynman diagrams based on the
`helicity amplitude approach', and the interesting reader may
consult Ref.\cite{BC1} for detailed calculation technology. The main
idea is to convert the problem into an equivalent `massless' one
that is well solved in literature, i.e. by transforming the massive
quark lines to be massless ones, and then to apply the symmetries as
much as possible. To extend the symmetries for the amplitude
corresponding to 36 Feynman diagrams, we first focus on the
numerator of the amplitude related to typical fermion lines, and
neither consider the color factors nor distinguish the flavor of the
fermion lines at the moment. Then, because of Feynman diagram
symmetries, these diagrams can be grouped into a few typical ones
according to the different type of fermion lines. And then we
implement proper factors for the fermion lines: color factors,
suitable denominator and spinors and $etc.$, so as to obtain an
exact and full typical fermion line that appears in Feynman
diagrams. When all kinds of typical fermion line factors, factors
for external lines of gluons and gluon propagators are `assembled',
then the full term, corresponding to the Feynman diagram of the
amplitude, is achieved. Next, to do the phase space integration, we
first use RAMBOS \cite{rkw} routine to generate the requested phase
space points and then use VEGAS \cite{gpl} program to perform the
integrations.

Based on the above calculation technology, we present the numerical
results. As for the present LO estimation, $f_{B_{s}}$ appears in
the amplitude as a linear factor, so the production cross sections
are proportional to it squared. Therefore, the uncertainties in the
production from $f_{B_s}$ can be figured out straightforwardly, so
throughout the paper, we take $f_{B_s}=0.209$ GeV\cite{QCDsr1}. And
because the spin splitting effects are ignored here, so there is no
difference for the decay constant between the spin stats $[^1S_0]$
and $[^3S_1]$. Further more, we shall study the uncertainties in `a
factorization way' throughout the paper, i.e., all of the parameters
vary independently in their reasonable regions. For instance, when
focussing on the uncertainties from the constitute $s$-quark mass
$m_s$, we let it be a basic `input' parameter varying in a possible
range
\begin{equation}
0.4 GeV \leq m_{s} \leq 0.7 GeV, \label{s-mass}
\end{equation}
with all the other factors, including the $B_s$-meson mass, the
decay constant  $f_{B_s}$ and {\it etc.} being fixed.

\begin{table}
\begin{center}
\caption{Total cross section for the hadronic production of
$B_s[1^{1}S_{0}]$ and $B_s^*[1^{3}S_{1}]$ with varying $m_s$, where
$m_b=4.9$ GeV, $m_{B_s}=m_s+m_b$, the gluon distribution function is
taken from CTEQ6L, $\mu_F^2=p_{T_{Bs}}^2+m_{Bs}^2$ and $\alpha_s$ is
of leading order.}\vskip 0.5cm
\begin{tabular}{|c||c|c|c|c||c|c|c|c|}
 \hline - &\multicolumn{4}{c||}{TEVATRON($\sqrt S=1.96$
TeV)}& \multicolumn{4}{c|}{LHC~($\sqrt S=14.$ TeV)}\\
\hline $m_s$ (GeV) & ~~~$0.4$~~~ & ~~~$0.5$~~~ &~~~$0.6$~~~&
 ~~~$0.7$~~~ & ~~~$0.4$~~~ & ~~~$0.5$~~~ & ~~~$0.6$~~~& ~~~$0.7$~~~\\
\hline $\sigma_{B_{s}}(nb)$ & 48.25 & 24.43 & 14.33 &
9.326 & 512.6 & 262.2 & 155.1 & 101.5  \\
\hline $\sigma_{B^{*}_s}(nb)$ & 165.5 & 82.18 & 46.78 & 29.30 &
1739. & 871.8 & 500.5& 316.2\\
\hline\hline
\end{tabular}
\label{tabms}
\end{center}
\end{table}

\begin{figure}
\centering
\hfill\includegraphics[width=0.48\textwidth]{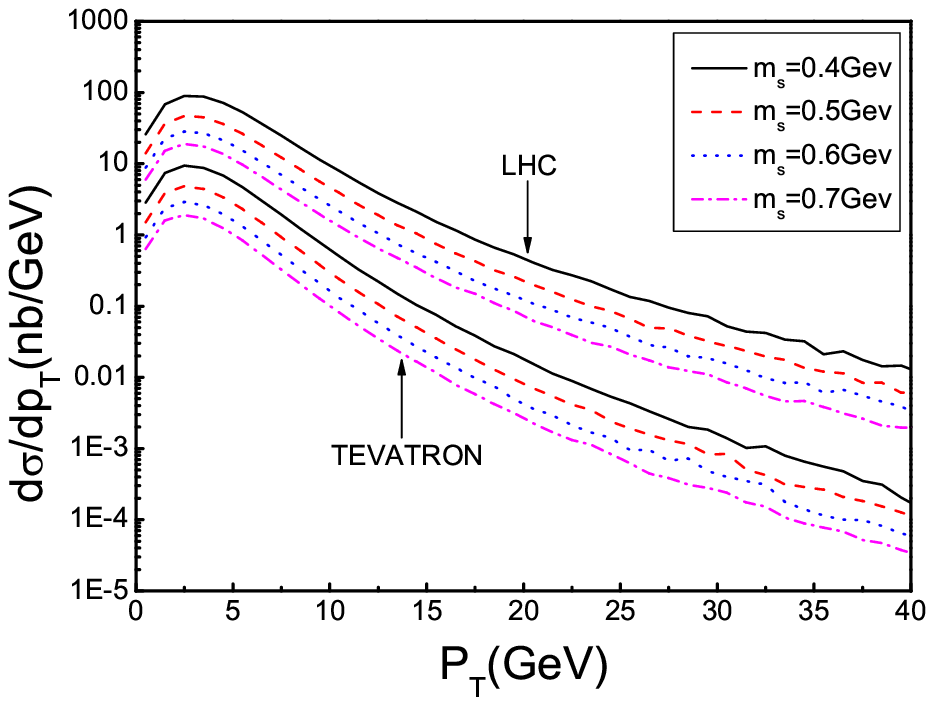}%
\includegraphics[width=0.48\textwidth]{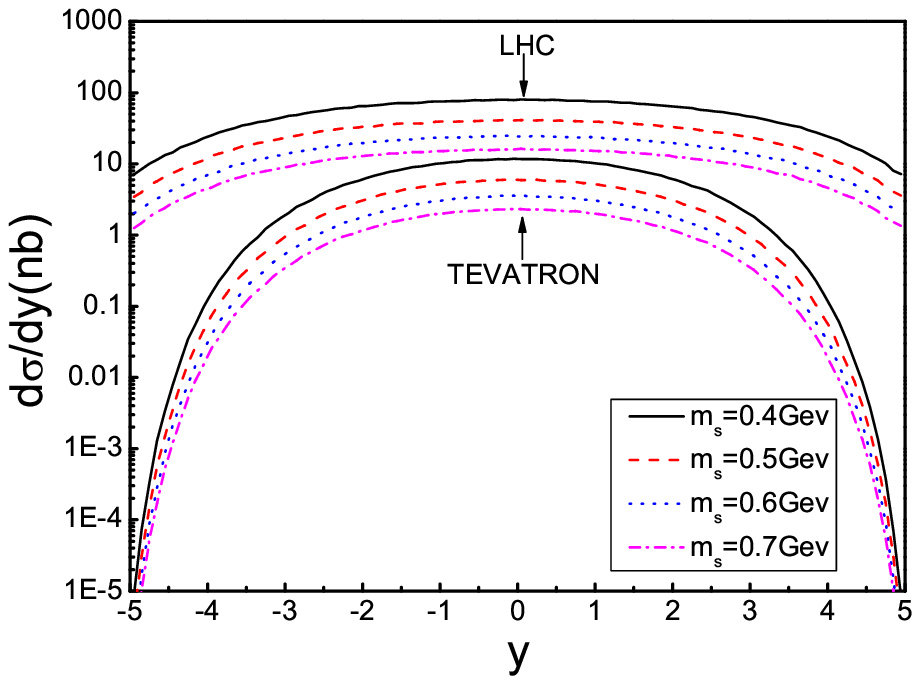}\hspace*{\fill}
\caption{$B_s$ differential distributions versus its $P_T$ and
rapidity $y$ with varying $m_s$. The gluon distribution function is
taken from CTEQ6L, the characteristic energy scale is taken as
$Q^{2}=p_{T_ {Bs}}^2+m_{Bs}^2$, and $\alpha_s$ is of leading order.
The solid line, dashed line, dotted line and dash-dot line stands
for $m_s=0.4$, $0.5$, $0.6$ and $0.7$ GeV respectively. The upper
(lower) four lines corresponding to the distributions at LHC
(TEVATRON) respectively.}\label{ptms}
\end{figure}

\begin{table}
\begin{center}
\caption{Total cross section for the hadronic production of
$B_s[1^{1}S_{0}]$ and $B_s^*[1^{3}S_{1}]$ with various $m_b$, where
$m_s$ is fixed to be $0.5$ GeV, the gluon distribution function is
taken from CTEQ6L, the factorization energy scale is chosen
$\mu_F^2=p_{T_ {Bs}}^2+m_{Bs}^2$ and $\alpha_s$ is of leading
order.} \vskip 0.5cm
\begin{tabular}{|c||c|c|c||c|c|c|}
 \hline - &\multicolumn{3}{c||}{TEVATRON($\sqrt S=1.96$
TeV)}& \multicolumn{3}{c|}{LHC~($\sqrt S=14.$ TeV)}\\
\hline $m_b$ (GeV) & ~~~~$4.8$~~~~ & ~~~~$4.9$~~~~ & ~~~~$5.0$~~~~ &
 ~~~~$4.8$~~~~ & ~~~~$4.9$~~~~ & ~~~~$5.0$~~~~\\
\hline $\sigma_{B_{s}}(nb)$ &22.60& 24.43 &26.49 &
 280.9& 262.2 &245.7  \\
\hline $\sigma_{B^{*}_s}(nb)$ &88.86 & 82.18 & 76.09 &
818.1 & 871.8 &929.3\\
\hline\hline
\end{tabular}
\label{tabmb}
\end{center}
\end{table}

In TAB.\ref{tabms}, we show the uncertainties from $m_s$, where the
other factors are fixed precisely as: $m_b=4.9$ GeV, the PDFs are
taking as CTEQ6L \cite{6lcteq}; the strong coupling $\alpha_s$ is in
LO and the factorization energy scale is taken to be
$\mu_F^2=p_{T_{Bs}}^2+m_{Bs}^2$. Note that for the mass of $B_s$,
the experimental result is $m_{B_s}=5.3663\pm 0.0006$ GeV
\cite{PDG}, while the prediction  by lattice QCD is about $5.37$ GeV
\cite{lattBs-mass}. Thus with $m_b=4.9$ GeV and $m_{B_s}=m_b+m_s$
\footnote{This relation should be satisfied according to the gauge
invariance of the hard subprocess. Further more, at the present, we
treat $B_s$ as non-relativistic and weak binding state, then at LO
the relative momentum between the constitute quarks can be
ignored.}, the obtained $m_{B_s}$ is in the region of theoretical
prediction as well as experimental measurement. In
Table~\ref{tabms}, the total cross-section for the hadronic
production of $B_s[1^{1}S_{0}]$ and $B_s^*[1^{3}S_{1}]$ at TEVATRON
and LHC are computed. From TAB.\ref{tabms}, one may observe that the
total cross section of $B_s^*[1^{3}S_{1}]$ is about 3 times bigger
that of $B_s[1^{1}S_{0}]$, which roughly agree with the naive
counting of spin states. $m_s$ affects the total cross section
greatly, e.g. when $m_s$ increases by steps of $0.1$ GeV, then the
cross section of $B_s$ or $B_s^*$ decreases by about $80\%-100\%$.
To show this point more clearly, we draw the $B_s$-$P_T$ and
rapidity $Y$ distributions with $m_s=0.4$, $0.5$, $0.6$ and $0.7$
GeV respectively in FIG.\ref{ptms}. This implies that the present
treatment of s-quark as heavy quark is reasonable but may be not too
accurate, and a more accurate one, e.g. by including proper
relativistic effects into the bound state, maybe improve the
estimation, which is out of the range of the present letter. A
similar calculation by varying $m_b$ within its reasonable region
but with fixed $m_s=0.5$ GeV, as shown by TAB.\ref{tabmb}, which
shows that when $m_b$ increases by steps of $0.1$ GeV, the cross
section of $B_s$ or $B_s^*$ changes slightly, which is around
$10\%$. More precise values of $m_b$ and $m_s$ from potential model
or lattice QCD can make the our estimations more reliable. In the
following parts of the paper when examining the uncertainties from
other factors, we shall always take the center values of $m_{s}=0.5$
GeV and $m_b=4.9$ GeV, for the quark masses.

\begin{table}
\begin{center}
\caption {Total cross-section for the hadronic production of
$B_s[1^{1}S_{0}]$ and $B_s^*[1^{3}S_{1}]$ at TEVATRON and at LHC
with LO running $\alpha_s$ and the characteristic energy scale Type
A: $Q^2=\hat{s}/4$; Type B: $Q^2=p_{T}^2+m_{B_s}^2$ and Type C:
$Q^2=p_{Tb}^2+m_b^2$.} \vskip 0.5cm
\begin{tabular}{|c|c|c|c|c||c|c|c|c|}
\hline - & \multicolumn{4}{|c||}{TEVATRON($\sqrt S=1.96$
TeV)}& \multicolumn{4}{c|}{LHC($\sqrt S=14.$ TeV)}\\
\hline -  & CTEQ6L & CTEQ6L &MRST2001 & CTEQ6L & CTEQ6L &CTEQ6L
&MRST2001L& CTEQ6L\\
 \hline $Q^2$  & Type A & \multicolumn{2}{|c|}{Type B}& Type C & Type A &
\multicolumn{2}{|c|}{Type B} & Type C\\
\hline $\sigma_{B_{s}}(nb)$ & 17.55 & 24.43& 20.82 & 25.01&
203.5 & 262.3 & 221.7& 264.1   \\
\hline $\sigma_{B^{*}_s}(nb)$  & 58.37 &82.19& 69.83 & 83.89 &
675.8 & 871.8 & 734.2& 876.5\\
\hline \hline
\end{tabular}
\label{pf}
\end{center}
\end{table}

\begin{figure}
\centering
\hfill\includegraphics[width=0.48\textwidth]{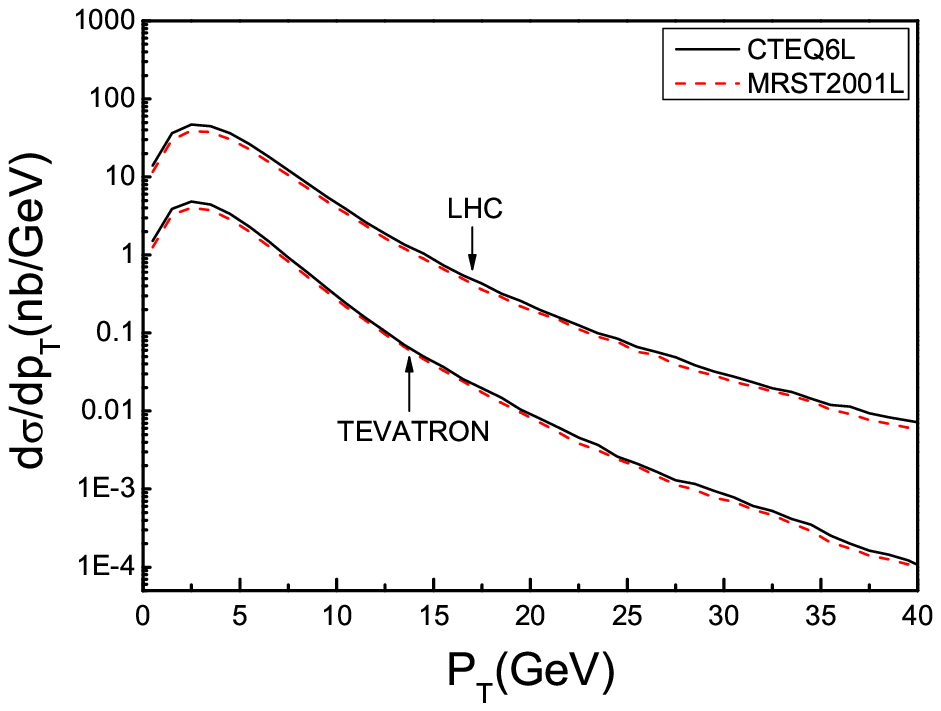}%
\includegraphics[width=0.48\textwidth]{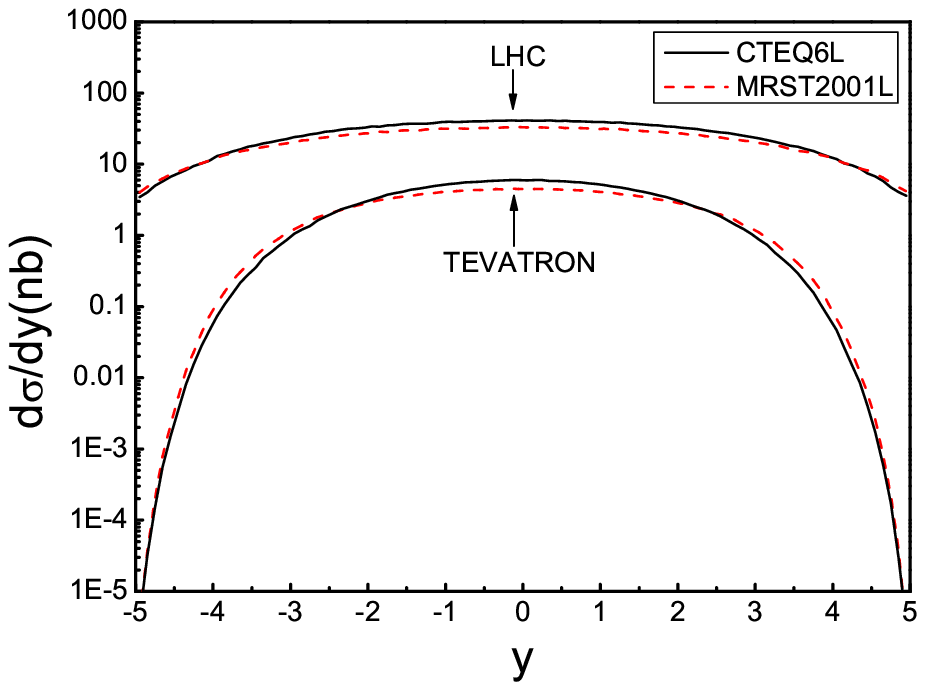}\hspace*{\fill}
\caption{$B_s$ differential distributions versus its transverse
momentum $P_T$ and rapidity $y$ for different LO PDFs, where the
solid line and the dashed line are for CTEQ6L and MRST2001L
respectively. The characteristic energy scale is taken as
$Q^2=p_{T}^2+m_{B_s}^2$. The upper and lower two lines corresponding
to the distributions in LHC and TEVATRON accordingly. }
\label{pdft}\vspace{-0mm}
\end{figure}

\begin{figure}
\centering
\hfill\includegraphics[width=0.48\textwidth]{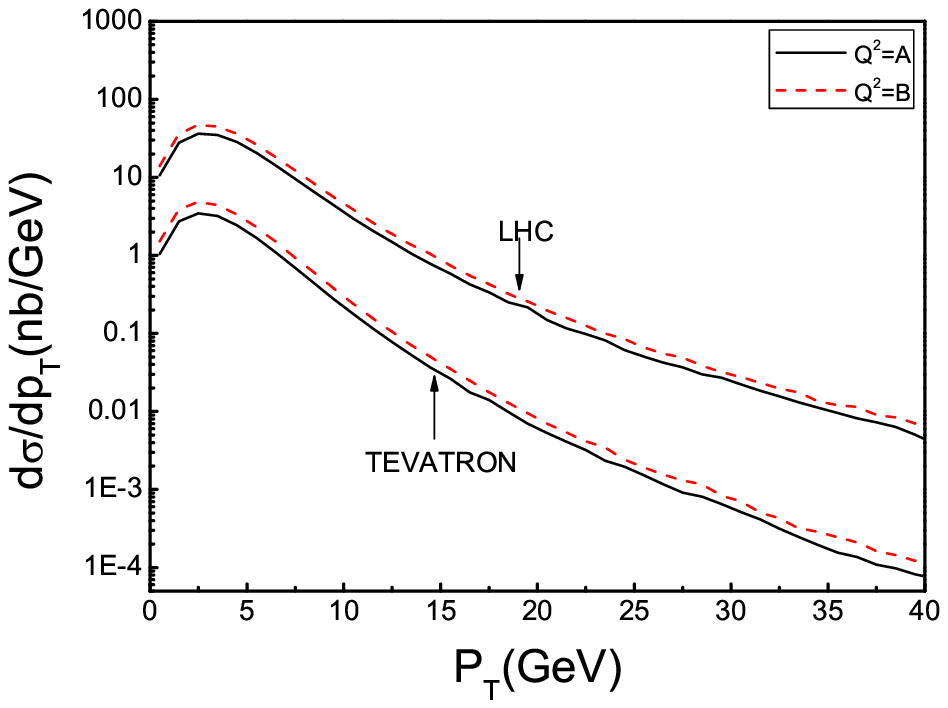}%
\includegraphics[width=0.48\textwidth]{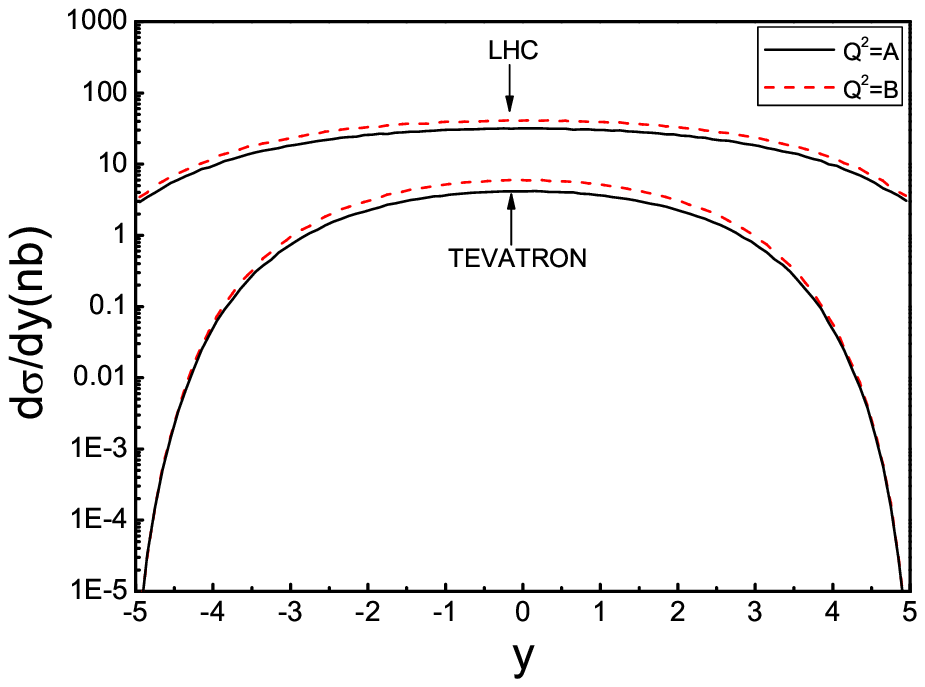}\hspace*{\fill}
\caption{$B_s$ differential distributions versus its transverse
momentum $p_T$ and rapidity $y$ for typical choices of $Q^2$, where
the solid and the dashed lines are for Type A and Type B
respectively. The gluon distribution is chosen as CTEQ6L and the
running $\alpha_s$ is in leading order. The upper and lower two
lines corresponding to the distributions in LHC and TEVATRON
accordingly. } \label{cteqQtpt}\vspace{-0mm}
\end{figure}

PDF is of non-perturbative nature, which can be obtained through
global fitting of the experimental data. Here we take CTEQ6L
\cite{6lcteq} and MRST2001L\cite{2001lmrst} as typical examples to
study the uncertainty caused by PDF. As shown in Eq.(\ref{cross})
PDF can be factorized out at the energy scale $\mu_F^2$ with the
help of pQCD factorization theorem. The factorization scale
$\mu_F^2$ can be usually taken as the characteristic energy scale
for the hard subprocess ($Q^2$), i.e. $\mu_F^2=Q^2$. For the present
case, the gluon-gluon fusion subprocess is of three-body final state
and contain heavy quarks, so there are ambiguities in choosing $Q^2$
and various choices of $Q^2$ would generate quite different results.
Since such kind of ambiguity cannot be justified by the LO
calculation itself, so we take it as the uncertainty of the LO
calculation. In the following we choose three typical examples to
study this kind of uncertainties: Type A: $Q^2=\hat{s}/4$, the C.M.
energy squared of the subprocess that is divided by $4$; Type B:
$Q^2=p_{T}^2+m_{B_s}^2$, the transverse mass squared of the $B_s$
meson; and Type C: $Q^2=p_{Tb}^2+m_b^2$, the transverse mass squared
of the $b$ quark. For comparison between TEVATRON and LHC and to
pinpoint the uncertainties from PDFs, $\alpha_s$ running and the
choices of the characteristic energy scale $Q^2$, we calculate the
production cross sections according to two types of PDFs, the strong
coupling $\alpha_s$ fixed by the corresponding PDFs and the
characteristic $Q^2$ chosen as Type A, Type B and Type C. The
obtained results are shown in TAB~\ref{pf}. From TABLE~\ref{pf}, it
is found that the difference caused by the two LO PDFs is small,
which is $\sim 15\%$. The choice of $Q^2$ as Type A and Type B cause
changes is somewhat larger, i.e. $20\%-30\%$, while the choose of
Type B and Type C leads to negligible changes to the cross section
(less than $1\%$). The total cross section of the $B^{(*)}_s$
production at LHC are at least one order larger in magnitude than
that at TEVATRON. This is mainly due to the fact that LHC ($\sqrt
S=14.$ TeV) has much higher collide energy than TEVATRON ($\sqrt
S=1.96$ TeV), so the lowest boundary of the gluon momentum fractions
$x_i \;\; (i=1,2)$ at LHC are much smaller than that at TEVATRON and
then there are more interacting gluons that have a C.M. energy above
the threshold for the subprocess, in the collision hadrons at LHC
than at TEVATRON. This can be shown more clearly by the $P_T$ and
$y$ differential cross sections. More explicitly, we draw the curves
for pseudo-scalar meson $B_s$ in FIGs.(\ref{pdft},\ref{cteqQtpt}).
FIG.\ref{pdft} shows that the differential distributions for the two
PDFs CTEQ6L and MRST2001L and FIG.\ref{cteqQtpt} shows that the
differential distributions for the two $Q^2$ Type A and Type B. In
regions of comparatively small $p_T$ and $|y|$, the distributions of
MSRT2001L are smaller than that of CTEQ6L. From the figure, we also
see that the $p_T$ distributions in TEVATRON are steeper than those
in LHC. From TABLE~\ref{pf}, we know that changes in the cross
section caused by Type B and Type C is quite small (less than
$1\%$), and the curves of the production obtained by Type B and Type
C are almost overlap, so we do not draw the curves for Type C.

\begin{table}
\begin{center}
\caption{Dependence of $R=\left(\frac{\sigma_{TEVTRON}}
{\sigma_{LHC}}\right)$ on $P_{Tcut}$ for $B_s[1^{1}S_{0}]$ and
$B_s^*[1^{3}S_{1}]$.} \vskip 0.6cm
\begin{tabular}{c|ccccc|cccccc}
\hline - & \multicolumn{5}{c|}{$B_s$} & \multicolumn{5}{c}{$B_s^*$}\\
\hline $P_{Tcut}(GeV)$ & ~~~0~~~ & ~~~5~~~ & ~~~20~~~ & ~~~35~~~&
~~~50~~& ~~~0~~~ & ~~~5~~~ & ~~~20~~~ & ~~~35~~~& ~~~50~~\\
$R (\times 10^{-2})$& 9.32 & 7.71 & 3.00 & 1.67 & 0.70 &
9.43 & 7.80 & 3.12& 1.85& 0.87 \\
\hline\hline
\end{tabular}
\label{ratio}
\end{center}
\end{table}

Experimentally, when the produced $B_s$ and $B^{*}_s$ mesons with a
small $P_T$ or a large rapidity $y$ are too close to the collision
beam, they cannot be measured, so only `detectable' events should be
taken into account, i.e. events with proper kinematic cuts on $P_T$
and $y$ should be properly set in the estimates. As a comparison, we
define a ratio $R=\left(\frac{\sigma_{TEVATRON}}
{\sigma_{LHC}}\right)_{P_{Tcut}}$ to show how $P_{Tcut}$ affects the
integrated cross sections at TEVATRON and LHC, and the results is
shown in TAB.\ref{ratio}. It can be found that without $P_{Tcut}$,
the integrated cross section at LHC is about one order higher than
that at TEVATRON, and the value of $R$ decreases greatly with the
increment of $P_{Tcut}$, at about $P_{Tcut}\simeq 45$ GeV, $R\sim
1$.

\begin{figure}
\centering
\hfill\includegraphics[width=0.48\textwidth]{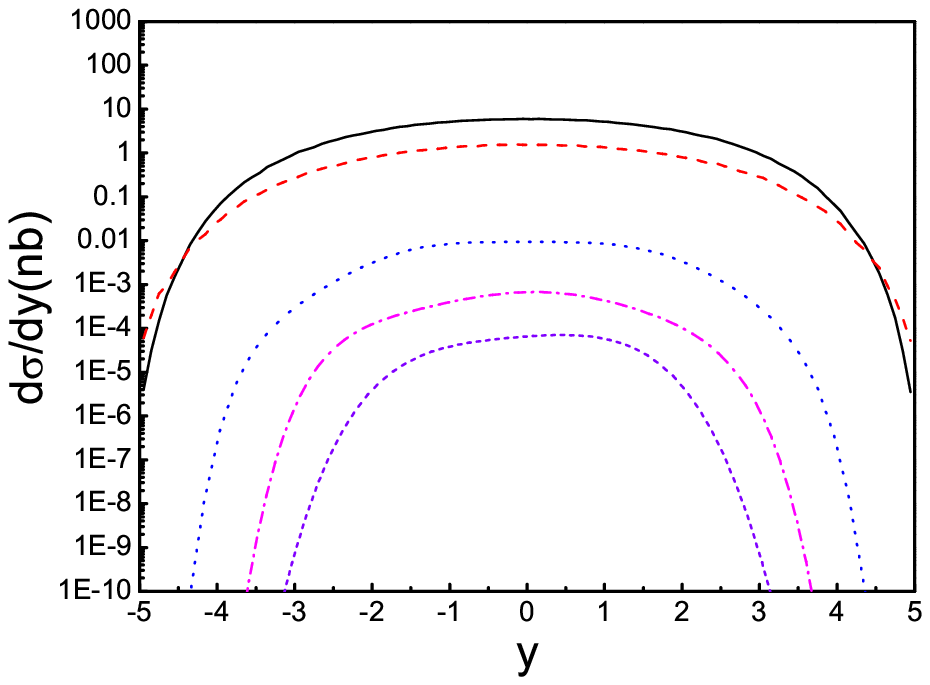}%
\includegraphics[width=0.48\textwidth]{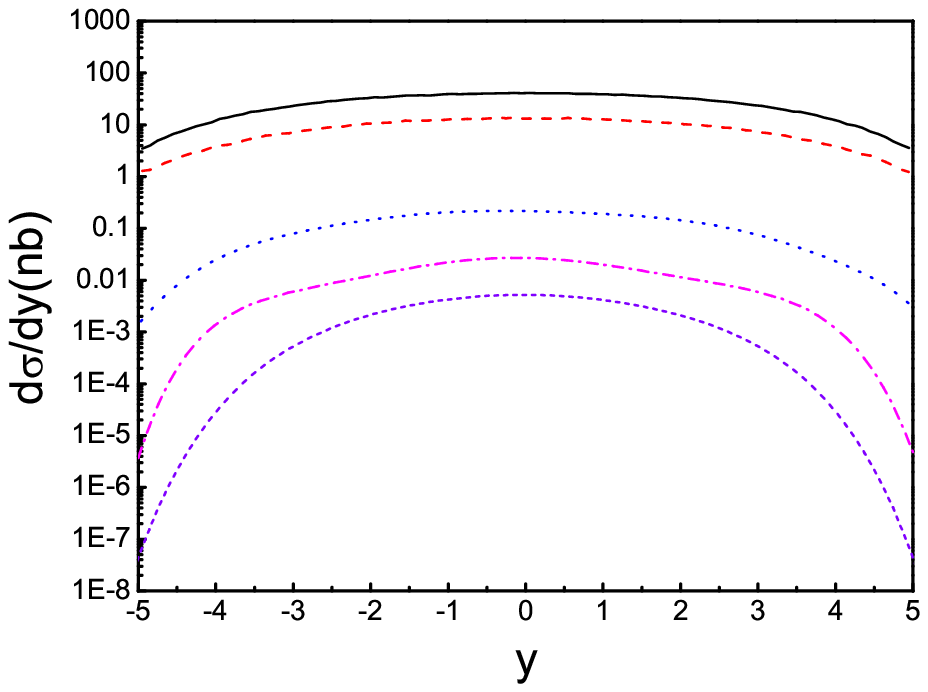}\hspace*{\fill}
\caption{$B_s$ differential distributions versus its $y$ with
various $p_{Tcut}$ in TEVATRON (left diagram) and in LHC (right
diagram). Solid line corresponds to the full production without
$p_{Tcut}$; dashed line to $p_{Tcut}=5.0$ GeV; dot line to
$p_{Tcut}=20.0$ GeV; the dash-dot line to $p_{Tcut}=35.0$ GeV; the
short dash line to $P_{Tcut}=50.0$ GeV .} \label{tptrap}
\end{figure}

\begin{table}
\begin{center}
\caption{Values of $R_{P_{Tcut}}$ for the hadronic production of
pseudo-scalar $B_s$ meson in TEVATRON and LHC.} \vskip 0.5cm
\begin{tabular}{|c||c|c|c||c|c|c||c|c|c||c|c|c||c|c|c|}
\hline\hline $p_{Tcut}$ & \multicolumn{3}{|c||}{$0.0$ GeV}&
\multicolumn{3}{|c||}{$5$ GeV}& \multicolumn{3}{|c||}{$20$ GeV}&
\multicolumn{3}{|c||}{$35$ GeV}
 & \multicolumn{3}{|c|}{$50$ GeV}\\
\hline $y_{cut}$& 1.0 & 1.5 & 2.0 & 1.0 &
1.5 & 2.0 & 1.0 & 1.5 & 2.0& 1.0 & 1.5 & 2.0& 1.0 & 1.5 & 2.0\\
\hline\hline $R_{p_{Tcut}}$ (TEVATRON) & 0.47 & 0.66  & 0.80 & 0.48
& 0.67 & 0.81 & 0.59 & 0.79 &
0.92 & 0.68 & 0.86 & 0.95 & 0.69 & 0.92 & 0.99\\
\hline $R_{p_{Tcut}}$ (LHC) & 0.32 & 0.47  &  0.60 & 0.33 & 0.48 &
0.61 & 0.40 & 0.57 &
0.71 & 0.45 & 0.62 & 0.77 & 0.56 & 0.73 & 0.83\\
\hline\hline
\end{tabular}
\label{tevcut}
\end{center}
\end{table}

Next, as an explicit example to show how the different cuts affect
the production, we study the distributions of $P_T$ and $y$ for
$B_s$. For the present purpose, we take CTEQ6L for PDF, LO running
$\alpha_s$ and Type B energy scale to carry out the study. The
correlations between $P_T$ and $y$ are interesting, so we plot the
$y$-distributions with various $P_T$-cuts over a wide range
$P_{Tcut}: 5.0\sim 35$ GeV in FIG.\ref{tptrap}. FIG.\ref{tptrap}
shows that the dependence of the differential distributions on
rapidity $y$ with different $P_{Tcut}$ at LHC exhibits a broader
profile than that at TEVATRON. The $p_T$-distributions of the
production vary with $y_{cut}$ mainly due to the fact that as $p_T$
increases, the dependence of the distribution on $y$ becomes smaller
as the value of $y_{cut}$ becomes less important. To analyze the
quantitative difference of the differential distributions with
regard to $P_{Tcut}$ and $y_{cut}$, we introduce a ratio for the
integrated hadronic cross sections, $ R_{P_{Tcut}}=
\left(\frac{\sigma_{y_{cut}}}{\sigma_{0}} \right)_{P_{Tcut}}$, where
$\sigma_{y_{cut}}$ and $\sigma_{0}$ are the hadronic cross section
with and without $y_{cut}$ respectively. The ratio $R_{p_{Tcut}}$
varies with $p_{Tcut}$ and $y_{cut}$, and its values are given in
TAB.\ref{tevcut}. TAB.\ref{tevcut} shows that for a fixed $y_{cut}$,
the value of $R_{P_{Tcut}}$ becomes larger with increasing
$P_{Tcut}$. It is understandable that the differential distributions
versus the rapidity $y$ decrease with the increment of $P_{T}$, so
the contributions to the hadronic cross section surviving after the
cut, i.e. $(|y|\leq y_{cut})$, increase with the increment of
$P_{Tcut}$.

To summarize: We have presented quantitative studies on the
uncertainties in estimates of the $B^{(*)}_s$ meson hadronic
production within the FFN scheme. The investigated quantitatively
uncertainties involve the PDF, the values of $m_b$ and $m_s$, and
the characteristic energy scale $Q^2$ of the process and etc.. It is
found that when $m_s$ increases by steps of $0.1$ GeV, the
integrated cross section of $B^{(*)}_s$ decreases by about
$80\%-100\%$, when $m_b$ increases by steps of $0.1$ GeV, it is
about $10\%$. While the uncertainties caused by the parton
distribution function and the factorization scale varies within the
region of $~\frac{1}{5}$ to $~\frac{1}{3}$. We have also shown the
differences between LHC and TEVATRON for various observable with
reasonable kinematic cuts, such as the cuts on the $B^{(*)}_s$ meson
transverse momentum $P_{Tcut}$ and rapidity $y_{cut}$. Our results
show that the experimental studies of the $B^{(*)}_s$ meson at the
two colliders are complimentary and stimulative. Concerning the
prospects of $B_s$ production at Fermilab TEVATRON and at CERN LHC,
the obtained results may be as useful references for these
experiments. Since LHC has much higher luminosity and higher
collision energy than that of TEVATRON, it seems that the
particularly interesting topics on $B_s$ may be more accessible and
fruitful at LHC than that at TEVATRON. Further more, it is
reasonable to assume that, similar to the hadronic production of
$J/\Psi$, $B_c$ and $\Xi_{cc}$ \cite{qiao,wu1,wu2}, the `heavy quark
mechanisms' via the sub-processes $g+s\to B^{(*)}_s+...$ and
$g+\bar{b}\to B^{(*)}_s+...$ may be as important as the gluon-gluon
fusion mechanism, which should be treated on the equal footing in
comparison to that of the gluon-gluon fusion mechanism. However to
be consistent theoretically and to deal with the possible double
counting from all these mechanisms, one should work in the
general-mass variable-flavor-number (GM-VFN) scheme
\cite{acot,gmvfn1,gmvfn2} in stead of the FFN scheme. A detailed
discussion on the GM-VFN scheme, and a comparison of $B^{(*)}_s$
production within these two schemes is in preparation and shall be
presented elsewhere.

\vspace{8mm}

{\bf Acknowledgments}: This work was supported in part by Natural
Science Foundation Project of CQ CSTC under Grant No.2008BB0298 and
Natural Science Foundation of China under Grant No.10805082 and
No.10875155, and by the grant from the Chinese Academy of
Engineering Physics under Grant No.2008T0401 and Grant No.2008T0402.

\end{document}